\documentclass[letter]{sigcomm-alternate} 
\usepackage{flushend}
\usepackage{array,multirow}
\usepackage{graphicx} 
\usepackage[usenames, dvipsnames]{color} 
\usepackage{enumitem} 
\usepackage[hyphens]{url}
\usepackage{subfigure}
\usepackage{amsmath}
\usepackage{amssymb}
\usepackage{cite}

\begin{document}

\title{A Primer on IPv4 Scarcity}

\numberofauthors{4}
\author{
\alignauthor\mbox{Philipp Richter}\\
    \affaddr{TU Berlin / ICSI}\\
   \affaddr{{prichter@icsi.berkeley.edu}}
\alignauthor \hspace{-0.7in}\mbox{Mark Allman}\\
    \affaddr \hspace{-0.7in}{ICSI}\\
   \affaddr \hspace{-0.7in}{{mallman@icir.org}}
\alignauthor \hspace{-1.4in}\mbox{Randy Bush}\\
    \affaddr \hspace{-1.4in}{Internet Initiative Japan}\\
    \affaddr \hspace{-1.4in}{{randy@psg.com}}
\alignauthor \hspace{-2in}\mbox{Vern Paxson}\\
    \affaddr \hspace{-2in}{UC Berkeley / ICSI }\\
    \affaddr \hspace{-2in}{{vern@cs.berkeley.edu}}
\vspace{1em}
\begin{tabular}{c}
\end{tabular}\\
\begin{tabular}{c}
\hspace{-7in}{\normalsize This article is an editorial note submitted to CCR. It has NOT been peer reviewed.}\\
\hspace{-7in}{\normalsize The authors take full responsibility for this article's
technical content. Comments can be posted through CCR Online.}
\end{tabular}
}

\maketitle

\begin{abstract}

With the ongoing exhaustion of free address pools at the registries serving
the global demand for IPv4 address space, scarcity has become reality.
Networks in need of address space can no longer get more address allocations 
from their respective registries.

In this work we frame the fundamentals of the IPv4 address exhaustion phenomena
and connected issues. We elaborate on how the current ecosystem of IPv4 address
space has evolved since the standardization of IPv4, leading to the rather
complex and opaque scenario we face today. We outline the evolution in
address space management as well as address space use patterns, identifying
key factors of the scarcity issues. We characterize the possible solution
space to overcome these issues and open the perspective of address blocks as
virtual resources, which involves issues such as differentiation between address
blocks, the need for resource certification, and issues arising when
transferring address space between networks.
 
\end{abstract}



\section{Introduction}
\label{sec:intro}

The Internet's design philosophy has facilitated enormous, rapid, and
de-centralized growth, from a specialized research facility to a massive
network of global importance.  In turn, the tremendous growth enabled by
the original design also outpaced engineers, researchers, and policy
makers.  This is clear in the numerous technical challenges that have
arisen---from the lack of support for traffic engineering and routing
security, to the scalability issues of the initial mapping of hostnames
to IP~addresses, to the lack of congestion control, to the inability to
accommodate mobility.

The community has largely been able to address such issues, albeit not
always in the most elegant way given that changing a running system
presents a challenging target in many cases.  However, under the
surface, policy and governance issues arose.  While scientists and
engineers often ignore these issues, they ultimately shape what we can
deploy in production.  In this paper we consider the entanglement of 
policies and governance with the technology for one of the Internet's
key resources: IP addresses.

Transmission of data between hosts across the Internet requires network
layer addresses to name the endpoints---i.e., IP addresses.  In IP
version~4, an address is represented by 32~bits in the IPv4 header;
hence there is a finite pool of roughly 4B addresses available.  The
network routing system cannot keep enough state to deal with each
individual address and therefore aggregates addresses into blocks.
Originally blocks were allocated quite informally, but as the Internet
grew the complexity of the process did as well.  At this point the
address space is nearly entirely allocated.

The Internet engineering community has long recognized the impending exhaustion
of IPv4, and in response designed a replacement network-layer protocol with
much longer addresses, IPv6.  However, given the network layer's critical
functionality, deploying IPv6 has proven difficult.\footnote{
	See \cite{ipv6sigcomm14} for an understanding of IPv6 deployment from 
	multiple viewpoints.
}  
Therefore, it is now widely acknowledged that IPv4 will continue to play a
significant role for a long time within the confines of the current resource
limits.  While there are technical maneuvers we can still make to cope---e.g.,
adding layers of network address translation (NATting)---IPv4 address blocks
have already become a good that people exchange on secondary markets.  This
reality brings yet another challenge to the Internet's policy and governance
ecosystem.

In this paper we first survey the evolution of the community's management of IP
addresses, for which we include both a discussion of the relevant policy
structures and organizations, as well as empirical illustrations of the
allocation and use of IP addresses over time.  This history then leads to a
set of observations about protocol design and the accompanying stewardship of
community resources.

\section{Evolution of Address Management}
\begin{figure} \centering
\includegraphics[width=0.95\linewidth]{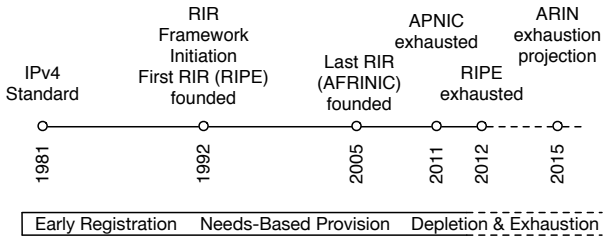} \caption{Evolution
of address management.} 
 \vspace{-1em}
\label{fig:rir_timeline} 
\end{figure}

From its standardization in 1981 \cite{rfc791} until now, the management
of IP addresses has undergone drastic change.  The changes were mainly a
result of the evolution of the Internet from a research network to a
global commercial network and the corresponding need to establish
international frameworks to manage its critical resources.  We
elaborate on this evolution in three time phases: The \textit{Early
  Registration} Phase starting with the arrival of IPv4, the
\textit{Needs-based Provision} Phase leading to the modern registry
framework, and the recently entered \textit{Depletion and Exhaustion} Phase.

\subsection{First Phase: Early Registration.}

Initially, address blocks were allocated quite informally, with Jon
Postel serving as the ``czar'' personally attending to each allocation.
Postel periodically re-published RFCs enumerating the current address
assignments (``please contact Jon to receive a number assignment'')
\cite{rfc790}. At that time, addresses block allocations came in one of
three \emph{classes}: class A networks ($2^{24}$ addresses), class B
($2^{16}$), and class C ($2^8$).  Classful addressing required a
\textit{network identifier} of one of these distinct types, meaning that an
operator requesting significantly more addresses than provided by a particular
threshold would instead be allocated a larger class network.  Given the
coarse-grained nature of the differences between these classes, this policy
led to heavy internal fragmentation and thus waste of address space.

Early (1981) in the Internet's evolution, parties had already registered
43~class~A networks, allocating in total more than 700M
addresses~\cite{rfc790}---vastly larger than the number of hosts actually
connected at that time.\footnote{
	Address registration statistics in terms of number of blocks and block 
	holders varied heavily among the first published RFCs.
}
While scarcity in address blocks was not mentioned as a looming issue, the
notion of different sizes of networks (A, B and C) suggests early
recognition of the finite nature of network address blocks and the need for
some sort of stewardship when parceling them out to different parties. The 
responsibility for the management of address space led to formalizing the 
notion of the IANA (first mentioned in IETF documents in 1990~\cite{rfc1060}),
and, in the same timeframe Solensky, drawing upon allocation statistics,
predicted IPv4 address exhaustion in the late '90s~\cite{solensky}.

\subsection{Second Phase: Needs-based Provision.}

The need for a more distributed and parsimonious framework to allocate IP
addresses---shaping the modern registry structure---appeared at least as
early as 1990 \cite{rfc1174}, with further refinements in 1992 and 1993
\cite{rfc1466}. The discussion at that time included the need to distribute
the administration of IP address blocks to regional registries, covering
distinct geographic regions to better serve the respective local
community---consciously fragmenting the registry. In addition, classless 
inter-domain routing (CIDR)\footnote{
  CIDR \cite{rfc1519} supported routing and forwarding on bit-aligned, as
  opposed to the previous byte-aligned, variable-length prefixes.  CIDR
  denotes prefixes as a combination of an IP address and a corresponding
  network mask, such as \textit{1.1.2.0/23} specifying a network with
  $2^{9}$ IP addresses that share their top 23~bits.  Introducing CIDR
  required significant network restructuring efforts, as well as changes to
  routing protocols and hardware (see, for example,
  \cite{Ford:1993:IRA:2329084.2329983}).
} and private address space\footnote{
	Reserved address blocks not
	globally routable, and thus usable
	concurrently within multiple networks as long as the given
	hosts do not require globally reachable IP~addresses~\cite{rfc1597}.
}
arose in 1993--4 to further conserve publicly routable address space.

The modern framework of Regional Internet Registries (RIRs), established in the years between 1992 and
2005, was very specific that conservation of address space was a primary
goal \cite{rfc2050}.  Five RIRs emerged, run as non-profit organizations:
RIPE for Europe in 1992, APNIC for the Asia-Pacific in 1993, ARIN for the
North-Americans in 1997, LACNIC for Latin America in 2002, and AfriNIC for
Africa in 2005.

The RIRs manage the distribution of IP address resources, each according to
their local policies. Policies within the RIRs are created using a community
process; for details of the process for each RIR, see
\cite{arin_policymaking,apnic_policymaking,ripe_policymaking,afrinic_policymaking,lacnic_policymaking}.
For the most part, anyone can submit an RIR policy proposal
which then undergoes an open discussion and review process, usually carried out
on mailing lists as well as in working group and policy meetings.
Adopting a proposal requires the community to reach a degree of consensus as
reflected in these discussions. 

\begin{figure} \centering
\includegraphics[width=0.85\linewidth]{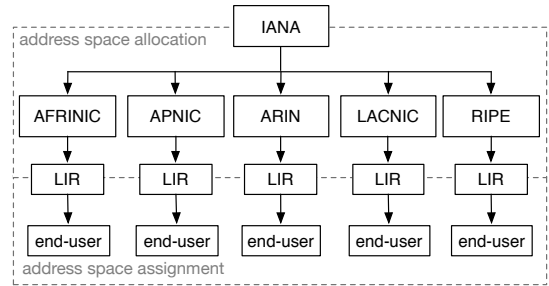}
\caption{Regional Internet Registry system.} 
\vspace{-1em}
\label{fig:rir_framework}
\end{figure}

We sketch the structure of the RIR framework in
Figure~\ref{fig:rir_framework}.  The IANA serves as the parent organization,
\textit{allocating} large free address blocks (/8, i.e. $2^{24}$ addresses,
granularity) to an RIR once their regional free pool reaches a low
threshold level.  The RIRs then further allocate subsets of these address
blocks to their members, the so-called LIRs (Local Internet Registries),
which are mainly ISPs. 
The LIRs then \textit{assign} address blocks to either smaller ISPs or 
for their own infrastructure. Thus, the allocation of a block  reserves it for 
(future) use, while the assignment of parts of an allocation puts that subset 
into use.\footnote{
	The APNIC and LACNIC regions also have \emph{National Internet  Registries} 
	(NIRs), which act as intermediaries between the RIR and the  LIR to serve 
	specific countries. For example, JPNIC does so for Japan.}
ISPs decide for themselves whether to become LIRs---meaning entering a direct
contractual relationship with the respective RIR---or to rely upon their
upstream provider to assign address space to them.\footnote{
	Under some circumstances, RIRs can also assign address space directly to 
	end users---so-called provider independent (PI) address space. Such
	assignments usually arise due to the user's
	need to connect to multiple upstream providers 
	(multi-homing), and thus requiring independent address space. For more 
	details, see for example \S~4.2 in the ARIN NRPM \cite{arin_manual} or the 
	RIPE policy documents \cite{ripe_policy}.  
	For a practical guide for operators, see \cite{van2002bgp}.
}

During the needs-based provision phase, one of the key principles was
that receivers of address space (LIRs) must \emph{justify} their need for
the address blocks they receive, though some RIRs no longer require this in
some contexts (e.g., RIPE for ``last /8'' allocations---see below).
LIRs requesting new allocations had to provide documentation showing a
sufficient \emph{utilization rate} of prior allocations, namely that a given
proportion of prior allocations were assigned to end-users, as well as
documentation of the intended use of new allocations. RIRs might also
request more detailed information, such as how many and what type of hosts
were connected to assigned subnets. LIRs passed these policies on to their
end-customers.  For example, if a customer of a transit provider required
blocks of IP addresses, they had to fill out corresponding LIR-specific forms
detailing the intended use of that block (e.g., \cite{ntt_policy}).

The global nature of the Internet raises the question of when an organization
is supposed to be served by a specific geographic region. Whether or not a
company can become an LIR under a specific RIR is not explicitly stated, but is
usually determined by the registered address of a company. However, there
also are organizations with multiple subsidiaries as members of---and holding
address resources from---multiple RIRs \cite{icann_legacy_meeting}. While 
address blocks are theoretically assigned and ``used'' by
organizations operating inside the region of the allocating RIR, current
policies are inconsistent regarding explicit constraints on the geographic 
region of an address block's actual use in the sense of where connected devices 
reside.\footnote{ARIN has a policy proposal to explicitly allow out-of-region
  use~\cite{arin_outofregion}, and a RIPE official stated that RIPE
  permits out-of-region use, assuming that the address blocks
  originate at some point from within the RIPE region (e.g., by a router
  at a European Internet Exchange
  Point)~\cite{ripe_outofregion}.   Numbering resources under the
  stewardship of LACNIC must be distributed among organizations legally
  constituted within its service region, and mainly serving networks and
  services operating in this region.  The AFRINIC community, on the
  other hand, has discussed explicitly limiting out-of-region use to
  prevent possible exploitation of their IP address resources from
  operators in other regions~\cite{afrinic_outofregion}.  }

\subsection{Third Phase: Depletion and Exhaustion.}

The five RIR communities agreed to a policy regarding address block allocation
upon the onset of exhaustion, which ICANN---the international body responsible
for the IANA function---ratified in 2009~\cite{icann_lastslash8}.
The policy dictated that when the IANA's IPv4 free pool 
reached five remaining /8 blocks, the IANA would distribute these blocks 
simultaneously and equally to the five RIRs.  In February 2011, the IANA allocated 
its last five free /8 address in accordance with the policy, one to each 
RIR~\cite{nro_depletion}. After that point, from a global perspective the pool 
of available IPv4 addresses was fully depleted.

\begin{figure}
  \includegraphics[width=0.9\linewidth]{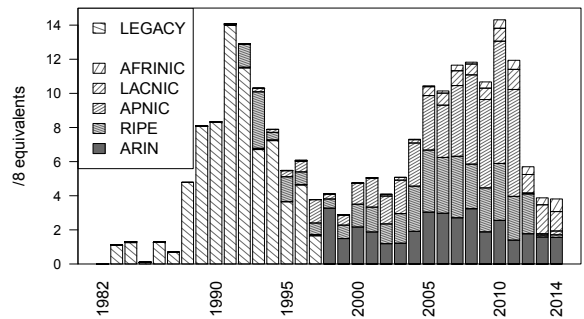}
  \caption{Yearly allocations of IPv4 address blocks.}
  \vspace{-1em}
  \label{fig:history_allocations}
\end{figure} 

Once the RIRs started to allocate from this last block from the IANA, the
``last~/8'' policies introduced by each RIR went into effect (e.g., APNIC's per
\cite{apnic_lastslash8}), imposing more restrictive allocation policies to 
further conserve this final address block and to allow new market entrants to 
still receive a last allocation, e.g., to implement IPv4-to-IPv6 transition 
mechanisms. Thus, LIRs could receive a single (small) allocation from this 
block. This transition occurred in April 2011 for APNIC, in September 2012 for 
RIPE, and in June 2014 for LACNIC, upon the exhaustion of their respective free pools. 
ARIN's exhaustion date is likely to occur in early 2015, while AFRINIC's pool 
should last until 2019~\cite{potaroo}.\footnote{
	We set the exhaustion date to when the RIRs started to
	allocate from their last /8, consistent with \cite{potaroo}.
}
LIRs in need of address space now need to find other means of obtaining
address space.

\section{Evolution of Address Block Allocation} 

Per the above, almost all of the free IP address blocks have
been distributed.  We can group today's address blocks into three
categories: (i)~blocks given out prior to the RIRs' existence, termed
\textit{legacy} address space;\footnote{LACNIC (and possibly AFRINIC)
  uses the date of ARIN's inception as their ``legacy'' threshold, not
  their own formation, as they would otherwise be unable to apply their
  policies to addresses that predate their formation.}
(ii)~blocks given
out during the era of the RIRs, termed \textit{allocated} address
blocks; and (iii)~\emph{reserved} address blocks, such as those set
aside for multicast and private addressing.

Figure \ref{fig:rir_timeline} on page~\pageref{fig:rir_timeline} shows a timeline of the most
significant events in the evolution of address block allocation.
One cannot pinpoint the transition between the above-mentioned phases 
precisely: the RIRs were founded years apart, hence ISPs 
in some regions received legacy address space for a longer period than in other
regions. ARIN, for example, began in 1997, whereas RIPE was founded in
1992. Thus, address space holders in the European region received allocated
address blocks earlier while holders in North America were still receiving legacy
blocks. The transition between phase~2 and~3 is ongoing, as two RIRs (ARIN and 
AFRINIC) still have unexhausted free pools.

In the remainder of this section we present an empirical lay-of-the-land
of the state of these allocations.

\subsection{History of Address Block Allocations}

The IPv4 address space consists of $2^{32}$ possible addresses, an equivalent
of 256 /8 address blocks. Of these 256 /8 blocks, 35.3 are reserved by the
IETF, e.g., for multicast, private use, and future use. This leaves
220.7 /8s worth routable address space.

In the following, we present a historical view on IPv4 address consumption 
from an RIR allocation point-of-view.  We rely on allocation files
provided by the RIRs~\cite{nro_files}. Figure~\ref{fig:history_allocations}
shows the address blocks given out by the registries over the years, as well as 
those given out prior to the existence of the modern RIR framework (shown as 
\textsc{legacy}).

Two peaks in address consumption are quite visible: The first 
occurs in the ``Early Registration Phase'' in the late 80's and early 90's. As
discussed in the previous section, address space conservation was not yet a
primary concern, and classful allocations resulted in heavy internal
fragmentation of address space.  The allocation rate drastically decreased in 
subsequent years, as address space conservation was implemented by the
RIRs. Address consumption rates in the late '90s and early 2000s suggested 
IPv4 address exhaustion would not happen  before 2020. The second peak, starting in 
the mid-2000s was dominated by allocations in the APNIC region, and comprised 
more than 50\% of all allocations in 2010 and 2011. After the 
exhaustion of the IANA free pool in 2011, a rapid decline in further allocations 
in 2012 is quite visible. Currently, fewer than 6~/8 equivalents are 
available for distribution by the RIRs.

The responsibility for the administration of legacy address blocks was 
transferred to ARIN upon its inception in 1997~\cite{nsi_arin_handover}. ARIN 
subsequently re-distributed some of these legacy blocks to the various other 
RIRs for respective holders located outside the ARIN region. This happened in 
the course of the ERX (Early Registrations Transfer) project~\cite{ripe_erx}. 
Yet, most legacy address space is still administered by ARIN, a symptom of North
America's dominance of the early Internet.

\begin{table}
\small
\begin{center}
\begin{tabular}{|l||r|r||r|}
\hline
  & \textbf{handed out} & \textbf{of which} & \textbf{available} \\ 
  & \textbf{/8s} & \textbf{legacy /8s} & \textbf{/8s} \\ 

\hline 
\hline 
ARIN & 100.5 & $\sim$ 64.9 & 0.35 \\ 
\hline 
RIPE & 47.6 & $\sim$ 11.93 & 0.97 \\ 
\hline 
APNIC & 51.0 & $\sim$ 4.40 & 0.74 \\ 
\hline 
LACNIC & 10.9 & $\sim$ 0.58 & 0.20 \\ 
\hline 
AFRINIC & 4.5 & $\sim$ 0.02 & 2.63 \\ 
\hline 
\hline
\textbf{total} & \textbf{214.5} & \textbf{$\sim$ 81.83} & \textbf{4.89} \\ 
\textbf{\textbf{\% of routable}} & \textbf{97.2\%} & $\sim$ \textbf{37.1\%} & 
\textbf{2.2\%} \\ 
\hline
\end{tabular} 
\end{center}
\caption{Address space statistics (February 2015).}
\label{tab:allocation_stats}
 \vspace{-1.5em}
\end{table}

Table \ref{tab:allocation_stats} gives an overview of the distribution of the
address space among the RIRs (in February 2015). The first column is the 
number of /8 equivalents, as listed in the allocation files of the RIRs.  The 
second column is an estimate of how much address space is legacy 
(given out in Phase~1) for each RIR.\footnote{
	For ARIN, we consider all address blocks handed out prior to
	December 1997 as legacy.  For the other RIRs, we consider all address
	blocks transferred as part of the ERX project as legacy, in addition to 
	blocks \textit{25/8}, \textit{51/8}, \textit{53/8} and \textit{57/8} for 
	RIPE and \textit{43/8} for APNIC. Some of these blocks may have been 
	voluntarily returned or otherwise changed their status.  Thus, 	the 
	number of legacy blocks only serves to give a sense of the landscape.
}
The last column shows the number of /8s per RIR that are available for
allocation. We observe that close to 97\% of the IPv4 address space has 
already been allocated, with less than 3\% available for further allocation.  
Some address blocks are in a reserved state (e.g., for temporal assignments for 
Internet experiments or conferences), and thus neither available nor 
handed-out. The heavy allocation rates in the last years prior to 
exhaustion mainly reflect heavy consumption in the APNIC region.  This could 
reflect a degree of hoarding, but might simply reflect booming Internet 
deployment in Asia.

\subsection{History of Routing}

In the last section we outlined how the management of IP addresses evolved over
time. From a pure allocation perspective, the address space is now close
to fully 
exhausted.  One important question is the degree to which
allocation reflects actual use.  We can consider this in two parts: (1)~the
degree to which elements of allocated blocks are routed, and thus potentially
in use; (2)~the degree to which addresses within routed blocks are in fact
used.  Here we assess the first of these, as we can much more readily obtain
insight into it (via the global routing table as publicly available from the
RouteViews project) than we can for the second consideration.

\begin{figure} \centering
\includegraphics[width=0.95\linewidth]{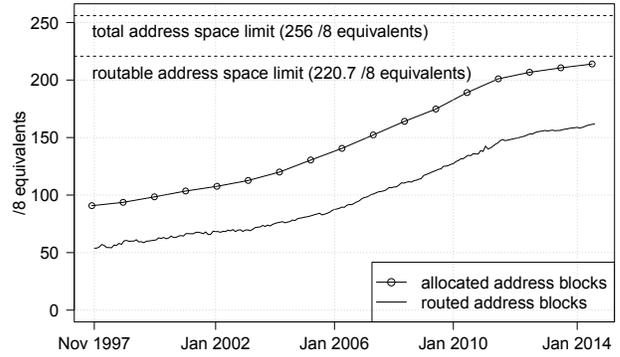}
\caption{Allocated and routed address blocks.}
 \vspace{-1em}
\label{fig:advertised_over_time_aggregate} 
\end{figure}

Figure \ref{fig:advertised_over_time_aggregate} shows the number of routed 
address blocks (expressed as /8 equivalents) over the last 16 years, along with 
the cumulative total of allocations made by the RIRs. We see that by 1997 
more than 25\% of the routable address space was advertised, which gradually 
increased to over 70\% in January 2014. While there is an increasing trend in 
the '00s, in the last two years the rate has been fairly stagnant,
perhaps reflecting address exhaustion. It should be noted  that the growth 
of the Internet in its early prime, starting in 1997,
used some 50\% of the available address space, while the 25\% routed prior
to that time
is likely due to classful allocations and rather lax allocation policies.

\begin{figure*} \centering
\includegraphics[scale=0.65]{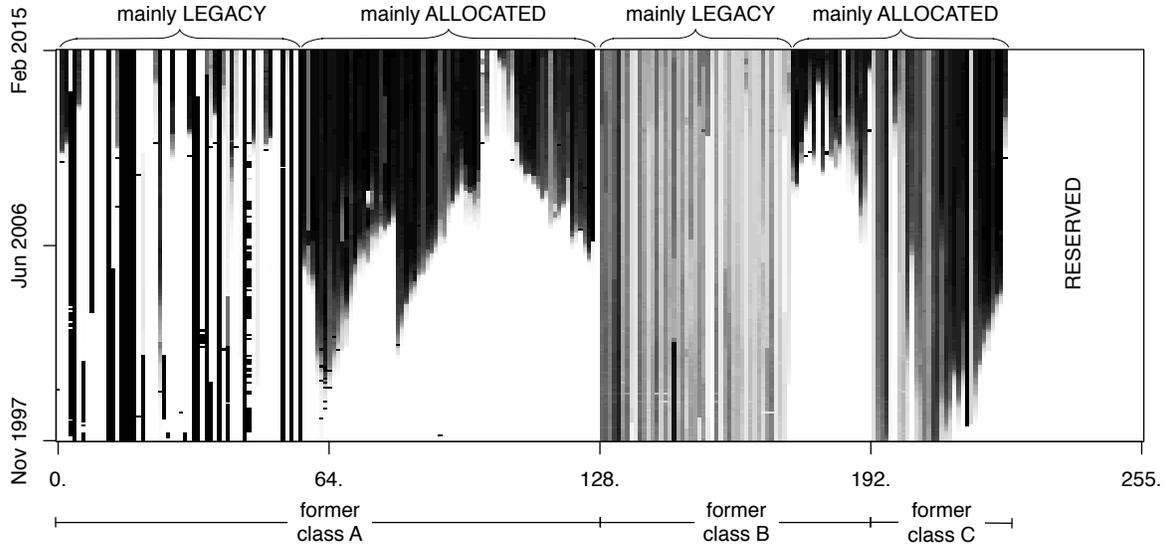}
\caption{Evolution of the distribution of routed address blocks
	over the total address space.}
 \vspace{-1em}
 \label{fig:advertised_over_time}
\end{figure*}

Figure~\ref{fig:advertised_over_time} shows the evolution of the routed address
space from 1997 until 2014, by plotting for each /8 the fraction of routed
addresses, ranging from white (no address blocks advertised) to black (all
address blocks advertised), with the various ranges annotated according to
their address types.\footnote{
	A few /8 \emph{legacy} block ranges of former class A networks
	were not given out, and are thus allocated.  In addition,
	some smaller address blocks in the former class B range were
	allocated by the RIRs, hence the notation ``mainly'' in the figure.}

The most striking observation from this plot is that the use of address
blocks is very unevenly distributed.  Address ranges assigned prior to
the existence of the RIRs, the legacy ranges, exhibit much fewer routed
address blocks, whereas the RIR-allocated ranges show a gradually increasing and
consistent routing pattern. We see that the measures taken in Phase~2, namely
the delegation of finer-grained address blocks (CIDR), together with the address
conservation principles of the RIRs, indeed had noticeable effect.
Hence, efficient address management greatly improved the utilization of address
space, but did not enhance utilization in legacy ranges outside of
their scope of operation. Today, address blocks in the legacy range have the 
greatest supply of free and usable address space. In fact, as of February 2015 
more than 90\% of the allocated address space is routed but only some 50\% of 
legacy address space.

The caveat when using routing tables to reason about the utilization of address
blocks is that, while it gives an indication of address space use (clearly
visible here), a routed address block does not necessarily mean that it is in
active use. Recent estimates range from 47\% to 60\%~\cite{dainotti_2014, 
lostinspace_2014, capturingghosts} of routed /24 address blocks that are 
actually used, meaning that they are actively engaged in communication. Actual 
use of address blocks can be measured actively (e.g., 
probing every IP address with a \textit{ping}) or by relying on passive 
measurements (e.g., identifying those parts of the address space that actively 
engage in communication---emitting traffic). Zander et 
al.~\cite{capturingghosts} also used meta-information such as Wikipedia edit 
logs and applied a statistical model in order to account for address blocks 
that are not detected by such methods.  While an address block 
being routed does not imply its actual use, unrouted address blocks, on the 
other hand, might be in private use for interconnecting networks not publicly 
reachable.

Hence, while the IP address space is close to fully exhausted, from an
allocation perspective, scarcity seems to be less of an issue from a
purely technical perspective (e.g., routing). While it requires further work to
quantify ``efficient use'', we can clearly see significant differences
between legacy address space and allocated address space.

\section{IP Addresses as a Resource}

IP addresses are virtual resources. In this section, we elaborate associated
issues.

\subsection{Addresses: All The Same, Only Different}

At first, one might consider IP addresses as a fully homogeneous
(fungible) resource, but in fact not all addresses have equivalent properties.
First, the size of a given address block governs its routability.  Larger
address blocks are less likely to be filtered by other operators, and can be
de-aggregated into smaller entities, allowing networks to better engineer their
route announcements.  In addition, an address block comes with \emph{history}:
for example, a block previously used by spammers will more likely be found on
blacklists, limiting one dimension of its usability. Finally, the properties of
address blocks differ depending on their allocation standing and any associated
policy restrictions, as noted in the next section.

\subsubsection{The case of allocated address blocks}

Allocated address blocks given out by the RIRs (Phase~2) are contractually
constrained---in a more or less explicit way---as not constituting the
\emph{property} of the respective holder. ARIN, RIPE and AFRINIC have explicit
``no property'' statements in the documents a receiver of address space must
agree to~\cite{arin_rsa,ripe_property,afrinic_property}, while LACNIC and APNIC
have more implicit statements in their contracts, not mentioning ownership or
property by name. LACNIC states that it can withdraw address blocks from
holders \cite{lacnic_property} and APNIC states that it [only]
hands out resources on a ``license basis''~\cite{apnic_property}. The RIRs apply
different policies for address space they give out, both with regard to
the requirement to document how address space is used as well as with regard
to transferability of address blocks. Hence, for RIR-allocated address blocks,
the holder will generally have to agree to policies and eventual policy changes
as imposed by the respective RIR. Thus, the region associated with an address
block directly affects the policies that govern it and thus also its value.

With respect to the possibility of RIRs unilaterally \textit{reclaiming} unused 
address space from LIRs, the policy documents differ. ARIN clearly rules out 
unilateral reclamation in its current RSA~\cite{arin_rsa}.
APNIC does not mention this possibility by name in its documents, but states
that \textit{``If an allocation or assignment becomes invalid then the address
space must be returned to the appropriate IR''}~\cite{apnic_property}. AFRINIC
states the possibility of \textit{``revocation or withholding of the service
supplied''}~\cite{afrinic_property}, and RIPE that it might deregister resources
if members fail to comply with their policies~\cite{ripe_property}.  We are not
aware of any cases of a unilateral reclamation of allocated address space to
date, aside from those where an address holder went defunct without successor.

\subsubsection{The case of legacy address blocks}

Legacy address blocks, on the other hand, are not in general governed by
contractual requirements imposed by any RIR. A noteworthy point with regard to
IP addresses as resources is the ongoing discussion whether IP addresses can be
considered property or not~\cite{rubi2011ipv4}.  Per
Figure~\ref{fig:advertised_over_time}, much of today's unrouted address space
is legacy, and thus not considered to be subject to current RIR policy.

The RIRs do maintain the registry databases and the anchors for reverse DNS
mappings for legacy blocks. However, the attitude of the RIRs towards holders
of legacy resources varies.  In the course of the last decades, the
RIRs---mainly ARIN~\cite{arin_reclaim}---started several initiatives to contact
holders of legacy address space with the goal of establishing some contractual
agreements between the holder and the RIR. As the documentation of legacy
allocations is often poor (e.g., outdated information), many holders of legacy
resources might not even be approachable. ARIN offers LRSAs (Legacy
Registration Services Agreement)~\cite{arin_lrsa} to holders of
legacy address space in their region. LRSAs establish a more formal
relationship between the address holder and ARIN, contain an explicit ``no
property'' clause, and  also contractually obligate the legacy holder
to ARIN's policies, including the policy for transfer to other entities (or when
the holder requests additional address space from ARIN).  In late 2007 ARIN
sent out more than 18K letters to legacy holders~\cite{arin_reclaim}.  Their
data shows that as of 3~years later, fewer than 1,000 LRSAs were in turn
requested by the holders,
and LRSAs cover less than 15\% of the legacy address space in 
the ARIN region~\cite{arin_lrsa_coverage}.
One address broker publicly suggests to legacy holders to not sign such 
LRSAs~\cite{donotsignrsa}. Another ARIN document states ``\textit{All 
of the IP address space that ARIN administers, including legacy space, is 
subject to ARIN policy}''~\cite{arin_transfer_legacy}. RIPE, on the other hand, 
adopted a proposal in February~2014 to offer registration services to holders 
of legacy address space and not impose particular regulations on transfers of 
registered legacy address blocks~\cite{ripe_legacy}.

Regarding the possibility of \textit{reclaiming} unused legacy address
blocks, ARIN states that it will not attempt to unilaterally reclaim
legacy address space~\cite{arin_lrsa_info}.  APNIC and RIPE ran
initiatives to contact holders of \textit{legacy} address blocks to recover 
address space \cite{ripe_reclaim, apnic_reclaim} but left
the decision up to the respective holder. In case of the RIPE
initiative, 400~holders were contacted of which 16~returned address
space to RIPE. However, there are prominent examples of voluntarily returned
legacy address blocks, such as Stanford University voluntarily
returning its /8 legacy address block in~2000~\cite{recovery_stanford}, as well 
as some other organizations~\cite{icann_recovery}.

A meeting convened by ICANN
in 2012 informally addressed issues related to legacy address
resources~\cite{icann_legacy_meeting}. The discussion involved representatives
from the RIRs, network operators holding legacy and non-legacy
address resources, and address brokers.  On one hand, it was argued that legacy resources by
their nature do not differ from other IP address blocks, and should thus be
subject to the same policies.  On the other hand, holders of legacy address
space argued that \textit{grandfathering} applies---meaning that as legacy
address space was given out prior to RIR policies, they are not subject to any
policies subsequently created by RIRs.

Hence, the open question with regard to legacy holders is whether they are
bound to the terms of the registry that currently provides registration
services to them---in a more general way, whether they hold ownership rights
for their addresses or not.

\subsection{Resource Certification \& Enforcement}

In the case of IP addresses, no global system exists to either authoritatively
verify the ownership of a given address block nor to prevent the usurping of
address blocks by illegitimate users. Inter-domain routing as instantiated by
BGP does not itself provide any mechanisms to ensure routing only by a block's
legitimate holder.  While the community readily recognizes BGP's lack of
security features, including its inability to authenticate routes, a large body
of research and accompanying deployment efforts has done little to change this
situation in productive environments (see \cite{butler2010survey} and
references therein).

The RIRs publicize the mapping of address spaces to their respective holders
via registry databases (WHOIS), which can be queried publicly, and by
delegating the respective reverse-DNS zones (\texttt{.in-addr.arpa}) to
authoritative nameservers specified by the address holders. This latter enables
the holders to specify PTR records for IP addresses in the respective namespace
(not a fundamental requirement or hallmark of ownership, but certainly
operationally useful).  Neither of these mechanisms provide sufficient
information to directly validate (or invalidate) route advertisements, such as
by authoritatively indicating the origin AS.\footnote{
	The ARIN WHOIS database recently started to provide a field for
	the origin AS, but the field is often unset and prominent cases
	of inconsistencies exist \cite{pch_origin}.
}
Thus, the administrative management of address space is largely decoupled from its actual
use.

The degree to which a prefix is usable by some entity---and which entities have
the capability to use it---simply depends on how far a route advertisement for
the given prefix propagates, which directly translates into how many hosts on
the Internet can interact with hosts in the given address block.

The propagation or non-propagation of prefix advertisements depends on the
route filtering performed by the border routers of ISPs.
To configure these filter settings, the community has established routing
registries (IRR), where network operators can register route objects to express
prefix ownership in the form of prefix-AS mappings \cite{butler2010survey}.
The various IRR databases are managed by several independent organizations,
including ISPs, RIRs and others \cite{irr_list}. However, not all address space
is registered in some registry (only around 50\% according to
\cite{irr50percent}) and information in these registries is known to  be
significantly inaccurate \cite{khan2013}. Many IRRs allow their participants to
introduce essentially any route object without further validation
\cite{bgpmon_rpki}. Complications with the IRR can again result in ISPs not
filtering advertisements from their peers using IRR information at all
\cite{deng2010evaluating}. 

There are well-known cases of erroneous IP address block advertisements, be it
hijacking of address blocks by spammers \cite{spam2006} or advertisements
caused by misconfigurations. As an example, a Pakistani ISP erroneously
advertised a prefix belonging to YouTube in~2008, resulting in an extensive
global outage for that service \cite{youtubehijack}.

The Internet Engineering Task Force (IETF) has developed a solution to this 
problem based on the RPKI (Resource Public Key Infrastructure)~\cite{rfc6480}. 
The basic function of the RPKI is to provide cryptographically verifiable 
attestations to address space and AS number allocations using a X.509 based 
hierarchy. RPKI uses the IANA and the RIRs as trust anchors, which give out 
certificates for resources they manage. Currently, RPKI services are offered by 
the RIRs as a free opt-in service only to their members.\footnote{ARIN requires 
legacy resource holders to sign an LRSA in order to be eligible to register 
their resources in the RPKI. Moreover, ARIN requires any operators wanting to 
\emph{use} the ARIN RPKI data to sign a Terms of Service Agreement that includes
an indemnification clause~\cite{arin_rpki_indemnity}. }

Based on the RPKI database, routers can verify that an AS advertising a 
specific 
prefix is in fact authorized to do so, which is referred to as RPKI-based 
origin validation~\cite{rfc7115}. This only prevents accidental advertisements 
and is not intended to prevent malicious attacks, as the full AS path is not
validated but only the origin of the path \cite{randy_rpki}.\footnote{To 
overcome this, AS-Path validation is necessary \cite{bgpsec}.} 
RPKI is supported by current routers from Cisco, Juniper, and Alcatel-Lucent, yet currently only 
around 5\% of the routable address space is covered by RPKI and by far the 
largest share of that address space is in the RIPE region 
\cite{ripe_rpki_stats}. 

While the problem of securing the advertisement of a prefix by only the
respective holder is well-known and many approaches have been proposed over the
years, little has changed in productive environments. 
Faced with the increasing scarcity of IP addresses (and the corresponding 
increasing value of addresses as resources), a functional scheme for certifying 
resources will be a key requirement in the near future in order to prevent 
illegitimate address space use.

\subsection{Address Markets} Given that free address pools are now
mostly exhausted and that demand for IPv4 address space will likely
continue to grow (at least until significantly broader IPv6 deployment),
address space transfers arise as a natural step necessary to further
distribute address space to those networks that need it. In light of the
issues discussed above---namely the fragmentation of addresses into
legacy and non-legacy address blocks subject to varying RIR-policies,
and connected ownership discussions, as well as the lack of widely
adopted resource certification mechanisms---the landscape of such
address transfers is at best murky.  Network operators have already
started buying and selling address blocks under varying conditions, as
we outline in the following.  This resulted in the emergence of several
\textit{address brokers} (e.g.,
\cite{addrex,iptrading,ipv4marketgroup}); companies that assist network
operators wanting to buy or sell address space.  Eventually, the RIRs
learned to encourage the use of address brokers to mediate transactions
within the strict confines of RIR policies.

\subsubsection{RIR Transfer Policies}

Today, four\footnote{
	AFRINIC states the possibility of transfers between LIRs 
	\cite{afrinic_policy}, but prohibits any such transfers 
	in their LSA unless they arise due to Mergers \& Acquisitions 
	\cite{afrinic_property}. 
} 
out of the five RIRs allow address space transfers among their members
\cite{ripe_policy,arin_manual,lacnic_policy,apnic_policy}.  In addition, ARIN,
RIPE and APNIC offer \textit{transfer listing services}
\cite{ripe_transfer_list,arin_transfer_list,apnic_transfer_list}, where network
operators can list address blocks they want to sell and express the need for
certain amounts of address space they want to buy.  These services aim to help
interested parties to come together, but use of them is not mandatory. RIPE
publicizes aggregated statistics for address space requests and offerings,
listing fewer than one million available addresses, and more than 17 million
requested addresses, as of February~2015 \cite{ripe_transfer_list}. These 
listings do not include any prices, as the negotiations remain entirely at the
discretion of the respective parties.

We observe a striking difference between how the RIRs perceive their roles when
it comes to conducting transfers under their policies. Except for RIPE, the
RIRs still require the receiving party of a transfer to \textit{justify} their
need for more address space according to their already established policies.
For example, APNIC requires transfer recipients to document use rates for past
allocations, as well as detailed plans for the use of transferred
resources~\cite{apnic_policy}, while ARIN states that recipients must
demonstrate the need for up to a 24-month supply following their established
policies~\cite{arin_manual}.

RIPE---as of February 2014---removed all ``justification of need'' clauses from
their policies. Address space can be transferred from any member to any
other member without the need to make statements of how the transferred
addresses will be used by the recipient. The proposal
\cite{ripe_reality} argued that address conservation will be in the interest of 
the members themselves (to not waste address space).  With regard
to concerns about possible address hoarding by wealthy LIRs it states that
``\textit{markets} [for other commodity goods] \textit{function well and in a
competitive manner, and there is no reason why the trade of IPv4 addresses will
be any different}''.

As of June 2014, \textit{Inter-RIR Transfers}---i.e., transfers between
address holders in different regions---are only possible between ARIN and APNIC. ARIN 
explicitly requires justification of need on the receiving side of a 
transfer---even if the recipient is located in a different region 
\cite{arin_interir}. Thus, RIPE's removal of justification of need in its 
transfer policies rules out Inter-RIR transfers with the ARIN region.

\begin{figure}

  \begin{center}
    \includegraphics[width=0.8\linewidth]{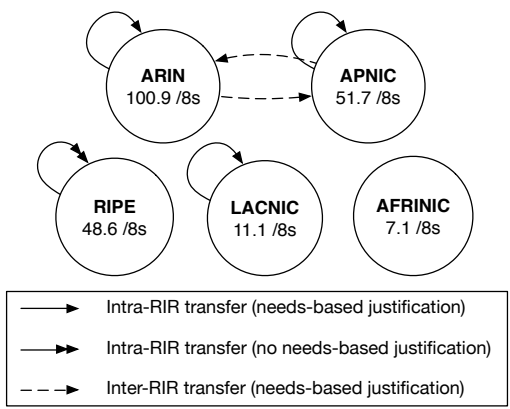}
  \caption{Current address space transfer policies of the RIRs and the   
  administered address space.}
   \vspace{-2em}
  \label{fig:rir_policies_bubble}
  \end{center}
    
\end{figure}

Figure \ref{fig:rir_policies_bubble} summarizes the transfer policies in place 
by the RIRs along with the address space they administer.

Another scenario in which address block transfers happen---and happened long
before modern transfer policies were established---is due to \emph{Mergers
\& Acquisitions}.  In this case, address blocks are part of the assets of a
company.  Since the related contracts are often confidential, these transfers
are not publicly listed by the RIRs---with the exception of APNIC, which 
requires full disclosure of the involved parties and publicly lists the 
corresponding address blocks~\cite{apnic_transfers}. The RIR's documents make no 
explicit statements about the justification of need for the transferred 
allocations. ARIN only states that the transferred resources will be subject to 
ARIN policies~\cite{arin_manual}, while APNIC states that it will ``review the 
status'' of the allocations, requiring full disclosure of all allocations held 
by the ``entities in question''. If that is not provided, APNIC will ``require 
that they be returned''~\cite{apnic_policy}.

\subsubsection{Transfers Outside the RIRs} 
Given that neither the legal nor the technical aspects of address space
transfers are under the full control of the current RIR framework, parties can
also conduct transfers separately from the RIRs.  To the extent that these
occur, a definitive determination of the party possessing a given allocation
becomes more difficult because the RIRs no longer possess accurate
records.

Even though ARIN states that legacy holders are subject to ARIN policies,
recent transfers, such as the well-known sale of more than 660K IP addresses
from the Nortel bankruptcy to Microsoft, have raised concerns whether they
complied with proper ARIN transfer policy.  Mueller et
al.~\cite{mueller2013dimensioning}
state that while ARIN was formally involved in
the transfer, likely no needs-based evaluation was performed on the receiving
side, and that ARIN's intervention boiled down to a \textit{``face-saving
exercise''}.  As the relationship of legacy holders
towards the RIRs is not entirely clear, one IP address trader has suggested
that legacy address holders in the ARIN region could de-register their address
space there and re-register it with a different RIR, such as
RIPE~\cite{donotsignrsa}. Doing so would effectively allow inter-region
transfers from ARIN to RIPE without undergoing any transfer process.
But currently there is no process to de-register from an RIR.

Aside from transfers that were formally noticed by the RIRs (such as the 
above example), address transfers can also happen without the
involvement of any registry at all. While address space can be of various types 
(\textit{legacy}, allocated to a holder by an RIR, assigned by a holder to an 
end-user, PI-assigned directly from the RIR to an end-user), bound to various 
contractual limitations, not much prevents any party from unofficially 
transferring an address block to another entity. This is known as a 
\textit{``black market''} transfer.  This possibility stems from the decoupled 
nature of address block management and actual address block use. If RIRs do not 
acknowledge such transfers, registry information becomes in turn inaccurate and 
incomplete, making the attribution of address blocks to their respective 
holders difficult.

In the simplest terms, we can view a transfer as simply an address block---or 
parts of it---formerly in use by some entity~A now being used by some 
entity~B, possibly outside the purview of any RIR regulation.  If the 
routing of the concerned address block is possible after
the transfer (it is not filtered by networks), and (to a lesser degree)
the corresponding reverse-DNS zones become under the control of the receiver
(e.g., by subdelegation of  reverse-DNS zones by the previous owner), the 
transfer would be successful.

It is unclear whether it is even feasible to detect the occurrence
of such transfers. Livadariu et al.\ attempted to detect such transfers by
looking for changes in routing origins over time~\cite{livadariu2013first}.
One difficulty here is that transferred address blocks are not necessarily
routed before they are transferred. Indeed, prior routing might be unlikely, as
unrouted address space is likely also unused and thus more likely to be
transferred. Also, whether such a transfer would be reflected in the
reverse-DNS is unclear, as NS records might simply not be changed and PTR
records might be unchanged or switched off.  Shifts in traffic, latency changes
or geographical changes might be due to transfers but also due to
restructurings within a company.
 
Thus, defining the boundaries of what exactly an address transfer is and what
it is not is not straightforward.  It is likely that the official RIR transfer
policies only cover a fraction of the total address transfers occurring in
various instantiations of the above scenarios. While transfers undergoing the
RIRs policies are publicly listed
\cite{arin_transfers,ripe_transfers,apnic_transfers} and quantifiable, the
number of address transfers outside this framework is unknown and requires
further research.

\section{Overcoming Scarcity}

IP address scarcity has become reality. That is, today only ARIN and
AFRINIC still hand out address space under regular conditions, while RIPE's,
APNIC's and LACNIC's pools have become exhausted, and they only hand out
one small
allocation from their last /8 to a requesting LIR. Comparing allocation rates
in 2014 to allocation rates in previous years, it is clear that the supply of
address blocks from the RIRs cannot satisfy the demand. Thus, address shortage
problems require other approaches.  Generally, we can consider three possible
solution spaces for this problem: (i)~develop more address space by
adopting IPv6, (ii)~multiplex current IPv4 address space using address
sharing techniques such as Carrier-grade NAT (CGN), and/or (iii)~more
efficiently use the current IPv4 address space.

\textbf{Develop more address space.}
The successor to IPv4, IPv6 \cite{rfc1883}, extends the routable address space
by orders of magnitude.  (Its design also aimed to address some other 
shortcomings of IPv4, such as support for mobility and extensibility.) It 
reflects the ultimate natural solution to the scarcity problem.  The RIRs 
advocate its use (e.g., RIPE hands out remaining IPv4 address blocks only to 
LIRs that have already received an IPv6 allocation~\cite{ripe_policy}), and the 
community has undertaken many other efforts to promote IPv6 adoption (e.g.,
\cite{ipv6launch}).  Nevertheless, the fraction of both IPv6-enabled networks
as well as native IPv6 traffic on the Internet remains comparably
small---adoption of IPv6 remains problematic 
and only slowly increases.\footnote{
	As of February 2015, Google reports some 4.5\% of clients accessing Google 
	to be IPv6 enabled, with adoption rates as high as 28\% in 
	Belgium, around 10 to 15\% in the US and Germany, and increasing support
	in other European countries. Nonetheless, the per-host
	adoption rate still ranges at or below 1\% for most countries, including
	China, India and Russia \cite{google_ipv6}.}
IPv6 is by itself not compatible with IPv4, and requires
complex transition mechanisms to ensure compatibility between the 
IPv4 and IPv6 Internet (e.g., \cite{rfc4213}).

\textbf{Multiplex address space.}
Alternatively, we can get by with many fewer addresses by multiplexing. 
Enterprise networks have long employed NAT to avoid having to allocate 
individual public IP addresses to every Internet-attached device.  Today, 
numerous approaches  to perform address sharing at scale are 
available---see \cite{address_sharing_ton} for a comprehensive
study---and are
already in use by several large ISPs.  While widespread use of NAT raises 
concerns about eroding end-to-end connectivity and semantics, as well as 
concerns by law enforcement agencies due to the erosion of attribution of IP 
addresses to end-users~\cite{cgn_fbi}, it poses fewer compatibility issues than 
IPv6 when employed for legacy network infrastructure.  According to 
\cite{cgn_deployment}, already more than 3\% of Internet users are behind CGNs, 
and Web hosting companies already employ heavy address sharing.

\textbf{Use address space more efficiently.}
As visible in Figure~\ref{fig:advertised_over_time_aggregate}, about a third of
all Internet address blocks remain unrouted, and thus not in (at least public)
use.  Moreover, even routed address space is not necessarily in active use. As
mentioned above, recent studies find utilization levels for the routed address
space at around 50\% to 
60\%~\cite{dainotti_2014,lostinspace_2014,capturingghosts}.  Hence, significant 
usable address space remains.  Making more efficient use of address space will 
require adapting address management policies, guidelines and technologies, 
including the
difficult (both technically and politically) problem of re-assigning already
allocated address blocks.

Network operators are currently adopting all of these options to varying
degrees: IPv6 adoption, CGN, and address transfers. We would
expect that cost will determine the manner and timeline when different
options predominate.  We should in turn find these costs reflected in the price
of IPv4 addresses as exchanged via secondary markets.

\section{Outlook}

IPv4 address scarcity is an issue requiring attention from the networking and
research community. Depending on the success of transitioning towards
widespread use of IPv6, we face a mid-to-long term scenario in which IPv4
addresses will have significantly more demand than supply. The key question is
for how long the cost of IPv4 addresses will be viewed as lower than the cost
of transitioning to IPv6 or using CGN.

That said, we note that while the limited IPv4 address space clearly will not
suffice in the longer term to provide every Internet device with its own
address, the current scarcity arises due to address management practices, and
not (yet) due to protocol limitations. Large fractions of the address space
remain unrouted, and of those address blocks that are routed, again only a
fraction is actually in use.

Just how to adapt the governance of the available address space to the current
situation remains a pressing question. While it is unclear whether IP address
block holders have ownership rights for their IP~addresses, secondary markets
already exist to facilitate their exchange.  However, the uncertainties
associated with address space transfers---both the legal status of legacy
address blocks and the varying policies among RIRs when it comes to such
transfers---will also complicate how pricing develops.  This, in turn, makes it
increasingly difficult for network operators to make decisions on 
which technology to adapt when.

As IPv4 addresses become an ever more scarce resource, increasing numbers of transfers, 
both inside the RIR framework as well as outside, are likely. 
As transfers outside the RIR framework can result in less accurate 
registration data provided by the RIRs---which in turn limits the possibility 
to use formal defenses such as RPKI-based origin validation---address block
hijacking events presumably will also increase. Viewing IP addresses as 
resources, other issues arise, such as resource certification and the exercise 
of control over ``who uses what address space''. As an inherently global 
resource, it is questionable whether the distributed registry framework can 
cope with the looming issues and provide sufficient resource liquidity.  Future 
scenarios for the management could include a more competitive environment among 
RIRs, or even a re-centralization of the registries.

From a research perspective, several issues arise: How to overcome scarcity
issues? What technologies will have what impact on the Internet? Will the 
community succeed in fully deploying IPv6 within the next decade,
or will we find ourselves 
stuck in a long-term situation in which IPv4, IPv6 and technologies like CGN 
operate in parallel?  What will be the corresponding impact on the Internet 
topology, its performance, and its reliability?

How to effectively deploy resource certification of address blocks and how to 
ensure routing only by the respective holder? How commonly do address transfers 
occur outside the RIR framework and with what sort of historical development 
and likely future trends?  What measurements could inform recommendations on 
how to govern the address space, in light of both IPv4 and IPv6 allocations? 
Did the creation of the distributed registry framework influence topological 
properties?  How should the RIRs agree on implementing consistent policies? 

We argue that the Internet community as a whole would greatly benefit from
empirical studies tackling the above questions, which will both aid network
operators with resolving business-critical decisions, as well as policy makers
as they adapt to this new landscape and work towards ensuring further
unhindered growth of the Internet. 

\section*{Acknowledgments}
\label{sec:acks}

This work was partially funded by NSF grant
CNS-1111672 and DHS/ARL contract W911NF-05-C-0013.

\bibliographystyle{plain}
\bibliography{ccr-paper621}

\begin{thebibliography}{10}

\bibitem{addrex}
{Addrex}.
\newblock {IPv4 Address Broker}.
\newblock \url{http://www.addrex.net}.

\bibitem{afrinic_property}
{AFRINIC}.
\newblock {AFRINIC Service Agreement 2013}.
\newblock \url{https://www.afrinic.net/en/services/rs/rsa}.

\bibitem{afrinic_policy}
AFRINIC.
\newblock {IPv4 Allocation Policy (AFPUB-2005-v4-001)}.
\newblock
  \url{http://www.afrinic.net/en/library/policies/126-afpub-2005-v4-001}.

\bibitem{afrinic_outofregion}
{AFRINIC}.
\newblock {Out-Of-Region Use of AFRINIC Internet Number Resources
  (AFPUB-2014-GEN-002-DRAFT-01)}.
\newblock
  \url{http://afrinic.net/en/community/policy-development/policy-proposals/1157-out-of-region-use-of-afrinic-internet-number-resources}.

\bibitem{afrinic_policymaking}
{AFRINIC}.
\newblock {Policy Development Process in the AFRINIC service region
  (AFPUB-2010-GEN-005)}.
\newblock
  \url{http://www.afrinic.net/en/community/policy-development/251-policy-development-process-in-the-afrinic-service-region-afpub-2010-gen-005}.

\bibitem{apnic_lastslash8}
{APNIC}.
\newblock {APNIC IPv4 Address Pool Reaches Final /8 (2011-04-15)}.
\newblock \url{http://www.apnic.net/publications/news/2011/final-8}.

\bibitem{apnic_transfers}
{APNIC}.
\newblock {IPv4 Address Transfer Logs}.
\newblock \url{ftp://ftp.apnic.net/public/transfers/apnic/}.

\bibitem{apnic_transfer_list}
{APNIC}.
\newblock {IPv4 Transfers Listing Service}.
\newblock
  \url{http://www.apnic.net/services/become-a-member/manage-your-membership/pre-approval/listing}.

\bibitem{apnic_policymaking}
APNIC.
\newblock {Policy development process (APNIC-111)}.
\newblock
  \url{http://www.apnic.net/publications/media-library/documents/policy-development/development-process}.

\bibitem{apnic_property}
{APNIC}.
\newblock {Policy environment for Internet number resource distribution in the
  Asia Pacific (APNIC-125, 2011-05-09)}.
\newblock \url{https://www.apnic.net/policy/policy-environment/text}.

\bibitem{apnic_reclaim}
{APNIC}.
\newblock {prop-017: Recovery of unused address space (2004-02-26)}.
\newblock \url{https://www.apnic.net/policy/proposals/prop-017}.

\bibitem{apnic_policy}
APNIC.
\newblock {Transfer, merger, acquisition, and takeover policy (APNIC-123)}.
\newblock \url{http://www.apnic.net/policy/transfer-policy}.

\bibitem{arin_policymaking}
{ARIN}.
\newblock {ARIN Policy Development Process}.
\newblock \url{https://www.arin.net/policy/pdp.html}.

\bibitem{arin_manual}
{ARIN}.
\newblock {ARIN's Number Resource Policy Manual (NRPM) Version 2014.02}.
\newblock \url{https://www.arin.net/policy/nrpm.html}.

\bibitem{arin_outofregion}
ARIN.
\newblock {Draft Policy ARIN-2014-1: Out of Region Use.}
\newblock \url{https://www.arin.net/policy/proposals/2014_1.html}.

\bibitem{arin_transfers}
{ARIN}.
\newblock {Inter-RIR and Specified Transfers of Internet Number Resources}.
\newblock \url{https://www.arin.net/knowledge/statistics/transfers.html}.

\bibitem{arin_interir}
ARIN.
\newblock {Inter-RIR Transfers.}
\newblock \url{https://www.arin.net/resources/request/transfers_8_4.html}.

\bibitem{arin_lrsa_coverage}
{ARIN}.
\newblock {Legacy Registration Service Agreement Statistics}.
\newblock \url{https://www.arin.net/knowledge/statistics/legacy.html}.

\bibitem{arin_lrsa_info}
{ARIN}.
\newblock {Legacy Registration Services Agreement.}
\newblock \url{https://www.arin.net/fees/agreements/legacy.html}.

\bibitem{arin_reclaim}
{ARIN}.
\newblock {Legacy Registration Services Agreement Outreach (2009-03-03)}.
\newblock \url{https://www.arin.net/resources/legacy/outreach.html}.

\bibitem{arin_lrsa}
{ARIN}.
\newblock {Legacy Registration Services Agreement v3.0}.
\newblock \url{https://www.arin.net/resources/agreements/legacy_rsa.pdf}.

\bibitem{arin_rsa}
{ARIN}.
\newblock {Registration Services Agreement v11.0}.
\newblock \url{https://www.arin.net/resources/agreements/rsa.pdf}.

\bibitem{arin_rpki_indemnity}
{ARIN}.
\newblock {RPKI Terms of Service Agreement}.
\newblock \url{https://www.arin.net/resources/rpki/tos.pdf}.

\bibitem{arin_transfer_list}
{ARIN}.
\newblock {Specified Transfer Listing Service}.
\newblock \url{https://www.arin.net/resources/transfer_listing/}.

\bibitem{arin_transfer_legacy}
ARIN.
\newblock {Transfers to Specified Recipients.}
\newblock \url{https://www.arin.net/resources/request/transfers_8_3.html}.

\bibitem{cgn_deployment}
I.~Van Beijnum.
\newblock {With the Americas running out of IPv4, it’s official: The Internet
  is full (2014-06-12)}.
\newblock
  \url{http://arstechnica.com/information-technology/2014/06/with-the-americas-running-out-of-ipv4-its-official-the-internet-is-full/}.

\bibitem{van2002bgp}
I.~Van Beijnum.
\newblock {\em BGP: Building reliable networks with the border gateway
  protocol}.
\newblock Page 61ff. O'Reilly Media, Inc., 2002.

\bibitem{rfc7115}
R.~Bush.
\newblock {Origin Validation Operation Based on the Resource Public Key
  Infrastructure (RPKI)}.
\newblock RFC 7115 (Best Current Practice), January 2014.

\bibitem{randy_rpki}
R.~Bush, R.~Austein, S.~Bellovin, and M.~Elkins.
\newblock {The RPKI \& Origin Validation}.
\newblock {NANOG 52, 2001}.

\bibitem{butler2010survey}
K.~Butler, T.~Farley, P.~McDaniel, and J.~Rexford.
\newblock {A Survey of BGP Security Issues and Solutions}.
\newblock {\em Proceedings of the IEEE}, 98(1), 2010.

\bibitem{rfc1174}
V.G. Cerf.
\newblock {{IAB Recommended Policy on Distributing Internet Identifier
  Assignment and IAB Recommended Policy Change to Internet "Connected"
  Status}}.
\newblock RFC 1174 (Informational), 1990.

\bibitem{ipv6sigcomm14}
J.~Czyz, M.~Allman, J.~Zhang, S.~Iekel-Johnson, E.~Osterweil, and M.~Bailey.
\newblock {From Infancy to Adolescence: A Decade-long Measurement of IPv6
  Adoption}.
\newblock In {\em ACM SIGCOMM}, 2014.

\bibitem{dainotti_2014}
A.~Dainotti, K.~Benson, A.~King, K.~Claffy, E.~Glatz, and X.~Dimitropoulos.
\newblock {Estimating Internet Address Space Usage Through Passive
  Measurements}.
\newblock {\em ACM CCR}, 44(1), 2014.

\bibitem{lostinspace_2014}
A.~{Dainotti et al.}
\newblock {Lost in Space: Improving Inference of IPv4 Address Space
  Utilization}.
\newblock Tech. rep., CAIDA, Oct 2014.
  \url{http://www.caida.org/publications/papers/2014/lost_in_space/}.

\bibitem{rfc1883}
S.~Deering and R.~Hinden.
\newblock {Internet Protocol, Version 6 (IPv6) Specification}.
\newblock RFC 1883 (Proposed Standard), 1995.
\newblock Obsoleted by RFC 2460.

\bibitem{deng2010evaluating}
W.~Deng, P.~Zhu, X.~Lu, and B.~Plattner.
\newblock {On Evaluating BGP Routing Stress Attack}.
\newblock {\em Journal of communications}, 5(1):13--22, 2010.

\bibitem{cgn_fbi}
{Donald Codling, FBI}.
\newblock {Attribution  Growing Challenges For LEAs}.
\newblock CWAG Conference of Western Attorneys General 2013
  \url{http://www.cwagweb.org/pdfs/2013/privacy_conference_ppts/cwag%20preso%2013.ppt}.

\bibitem{Ford:1993:IRA:2329084.2329983}
P.~Ford, Y.~Rekhter, and H.~Braun.
\newblock {Improving the Routing and Addressing of IP}.
\newblock {\em IEEE Network}, 7(3), 1993.

\bibitem{rfc1519}
V.~Fuller, T.~Li, J.~Yu, and K.~Varadhan.
\newblock {Classless Inter-Domain Routing (CIDR): an Address Assignment and
  Aggregation Strategy}.
\newblock RFC 1519 (Proposed Standard), 1993.
\newblock Obsoleted by RFC 4632.

\bibitem{potaroo}
{Geoff Houston}.
\newblock {IP Address Report. \url{http://www.potaroo.net}}.

\bibitem{rfc1466}
E.~Gerich.
\newblock {Guidelines for Management of IP Address Space}.
\newblock RFC 1466 (Informational), 1993.
\newblock Obsoleted by RFC 2050.

\bibitem{google_ipv6}
{Google}.
\newblock {IPv6 Statistics.}
\newblock \url{https://www.google.com/intl/en/ipv6/statistics.html}.

\bibitem{rfc2050}
K.~Hubbard, M.~Kosters, D.~Conrad, D.~Karrenberg, and J.~Postel.
\newblock {Internet Registry IP Allocation Guidelines}.
\newblock RFC 2050 (Historic), 1996.
\newblock Obsoleted by RFC 7020.

\bibitem{icann_lastslash8}
{ICANN}.
\newblock {Global Policy for the Allocation of the Remaining IPv4 Address Space
  (ratified in 2009)}.
\newblock
  \url{https://www.icann.org/resources/pages/remaining-ipv4-2012-02-25-en}.

\bibitem{icann_legacy_meeting}
{ICANN}.
\newblock {Multi-Stakeholder Discussion: Legacy Internet Protocol (IP) Numbers
  in the Current Policy Environment (2012).}
\newblock Audio Transcript. \url{http://toronto45.icann.org/node/34377}.

\bibitem{icann_recovery}
{ICANN Blog}.
\newblock {Recovering IPv4 Address Space}.
\newblock \url{http://blog.icann.org/2008/02/recovering-ipv4-address-space/}.

\bibitem{ipv6launch}
{Internet Society}.
\newblock {World IPv6 Launch}.
\newblock \url{http://www.worldipv6launch.org}.

\bibitem{iptrading}
{IP Trading}.
\newblock {IPv4 Address Broker}.
\newblock \url{http://www.iptrading.com}.

\bibitem{ipv4marketgroup}
{IPv4 Market Group}.
\newblock {IPv4 Address Broker}.
\newblock \url{http://www.ipv4marketgroup.com}.

\bibitem{donotsignrsa}
{IPv4 Market Group Blog}.
\newblock {Don't Sign an RSA During Your 8.2 IPv4 Transfer.}
\newblock \url{http://ipv4marketgroup.com/?s=do+not+sign+rsa}.

\bibitem{irr_list}
{IRR}.
\newblock {List Of Routing Registries}.
\newblock \url{http://www.irr.net/docs/list.html}.

\bibitem{khan2013}
A.~Khan, K.~Hyon-chul, T.~Kwon, and Y.~Choi.
\newblock {A comparative Study on IP Prefixes and their Origin ASes in BGP and
  the IRR}.
\newblock {\em ACM CCR}, 43(3), 2013.

\bibitem{lacnic_policymaking}
{LACNIC}.
\newblock {Policy Development Process (version 2.0 - 01/09/2008)}.
\newblock
  \url{http://www.lacnic.net/en/web/lacnic/proceso-de-desarrollo-de-politicas}.

\bibitem{lacnic_policy}
LACNIC.
\newblock {Policy Manual v2.1 (2014-03-25)}.
\newblock \url{http://www.lacnic.net/web/lacnic/manual-2}.

\bibitem{lacnic_property}
{LACNIC}.
\newblock {Registration Services Agreement}.
\newblock \url{http://lacnic.net/docs/rsa-en.pdf}.

\bibitem{rfc6480}
M.~Lepinski and S.~Kent.
\newblock {An Infrastructure to Support Secure Internet Routing}.
\newblock RFC 6480 (Informational), February 2012.

\bibitem{bgpsec}
M.~Lepinski and S.~Turner.
\newblock {{An Overview of BGPSEC}}.
\newblock IETF Internet-Draft draft-ietf-sidr-bgpsec-overview-05
  \url{https://tools.ietf.org/html/draft-ietf-sidr-bgpsec-overview-05}, 2014.

\bibitem{livadariu2013first}
I.~Livadariu, A.~Elmokashfi, A.~Dhamdhere, and K.~Claffy.
\newblock {A First Look at IPv4 Transfer Markets}.
\newblock In {\em ACM CoNEXT}, 2013.

\bibitem{mueller2013dimensioning}
M.~Mueller, B.~Kuerbis, and H.~Asghari.
\newblock {Dimensioning the Elephant: An Empirical Analysis of the IPv4 Number
  Market}.
\newblock {\em info}, 15(6):6--18, 2013.

\bibitem{nsi_arin_handover}
{National Science Foundation (NSF)}.
\newblock {Amendment 7 to Cooperative Agreement Between NSI and U.S. Government
  (1997-12-03)}.
\newblock \url{http://archive.icann.org/en/nsi/coopagmt-amend7-03dec97.htm}.

\bibitem{youtubehijack}
RIPE NCC.
\newblock {YouTube Hijacking: A RIPE NCC RIS case study}.
\newblock
  \url{http://www.ripe.net/internet-coordination/news/industry-developments/youtube-hijacking-a-ripe-ncc-ris-case-study}.

\bibitem{recovery_stanford}
{Network World, Vol. 17, No. 4.}
\newblock {Stanford move rekindles 'Net address debate (2000-01-24)}.

\bibitem{rfc4213}
E.~Nordmark and R.~Gilligan.
\newblock {Basic Transition Mechanisms for IPv6 Hosts and Routers}.
\newblock RFC 4213 (Proposed Standard), October 2005.

\bibitem{nro_files}
{NRO}.
\newblock {Extended Allocation and Assignment Reports}.
\newblock \url{http://www.nro.net/statistics}.

\bibitem{nro_depletion}
{NRO}.
\newblock {Free Pool of IPv4 Address Space Depleted.}
\newblock \url{http://www.nro.net/news/ipv4-free-pool-depleted}.

\bibitem{ntt_policy}
{NTT America}.
\newblock {Request for IPv4 and IPv6 Address Space v3.7}.
\newblock {\url{https://whois.gin.ntt.net/ipj.txt}}.

\bibitem{pch_origin}
{Packet Clearing House (PCH)}.
\newblock {Routing Origin Inconsistency.}
\newblock
  \url{https://prefix.pch.net/applications/routing-origin-inconsistency/}.

\bibitem{rfc790}
J.~Postel.
\newblock {Assigned numbers}.
\newblock RFC 790 (Historic), 1981.
\newblock Obsoleted by RFC 820.

\bibitem{rfc791}
J.~Postel.
\newblock {Internet Protocol}.
\newblock RFC 791 (Internet Standard), 1981.
\newblock Updated by RFCs 1349, 2474, 6864.

\bibitem{spam2006}
A.~Ramachandran and N.~Feamster.
\newblock {Understanding the Network-level Behavior of Spammers}.
\newblock {\em ACM CCR}, 36(4), 2006.

\bibitem{rfc1597}
Y.~Rekhter, B.~Moskowitz, D.~Karrenberg, and G.~de~Groot.
\newblock {Address Allocation for Private Internets}.
\newblock RFC 1597 (Informational), March 1994.
\newblock Obsoleted by RFC 1918.

\bibitem{rfc1060}
J.K. Reynolds and J.~Postel.
\newblock {Assigned numbers}.
\newblock RFC 1060 (Historic), 1990.

\bibitem{ripe_erx}
{RIPE NCC}.
\newblock {Early Registration Transfer (ERX) Project}.
\newblock \url{http://www.ripe.net/lir-services/resource-management/erx}.

\bibitem{ripe_outofregion}
{RIPE NCC}.
\newblock {Ingrid Wijte on the RIPE policy mailing list. (2012-05-23)}.
\newblock
  \url{http://www.ripe.net/ripe/mail/archives/address-policy-wg/2012-May/006981.html}.

\bibitem{ripe_transfer_list}
{RIPE NCC}.
\newblock {IPv4 Transfer Listing Service}.
\newblock \url{http://www.ripe.net/lir-services/resource-management/listing}.

\bibitem{ripe_transfers}
{RIPE NCC}.
\newblock {IPv4 Transfer Statistics}.
\newblock
  \url{http://www.ripe.net/lir-services/resource-management/ipv4-transfers/table-of-transfers}.

\bibitem{ripe_policymaking}
{RIPE NCC}.
\newblock {Policy Development Process in RIPE (ripe-614)}.
\newblock \url{http://www.ripe.net/ripe/docs/ripe-614}.

\bibitem{ripe_property}
{RIPE NCC}.
\newblock {RIPE NCC Standard Service Agreement (ripe-533)}.
\newblock \url{http://www.ripe.net/ripe/docs/ripe-533}.

\bibitem{ripe_rpki_stats}
{RIPE NCC}.
\newblock {RPKI Statistics}.
\newblock \url{http://certification-stats.ripe.net}.

\bibitem{ripe_legacy}
{RIPE NCC}.
\newblock {Services to Legacy Internet Resource Holders (ripe-605)}.
\newblock \url{http://www.ripe.net/ripe/docs/ripe-605}.

\bibitem{ripe_reclaim}
{RIPE NCC}.
\newblock {Status of Legacy IPv4 Address Space (2011-09-12)}.
\newblock
  \url{https://labs.ripe.net/Members/xavier/status-of-legacy-ipv4-address-space}.

\bibitem{ripe_reality}
{RIPE NCC}.
\newblock {Post Depletion Adjustment of Procedures to Match Policy Objectives,
  and Clean-up of Obsolete Policy Text}.
\newblock \url{http://www.ripe.net/ripe/policies/proposals/2013-03}, 2013.

\bibitem{ripe_policy}
{RIPE NCC}.
\newblock {IPv4 Address Allocation and Assignment Policies for the RIPE NCC
  Service Region (ripe-606)}.
\newblock \url{http://www.ripe.net/ripe/docs/ripe-606}, 2014.

\bibitem{rubi2011ipv4}
E.~M. Rubi.
\newblock {The IPv4 Number Crisis: The Question of Property Rights in Legacy
  and Non-Legacy IPv4 Numbers}.
\newblock {\em American Intellectual Property Law Association (AIPLA) Quarterly
  Journal}, 39:477, 2011.

\bibitem{address_sharing_ton}
N.~Skoberne, O.~Maennel, I.~Phillips, R.~Bush, J.~Zorz, and M.~Ciglaric.
\newblock {IPv4 Address Sharing Mechanism Classification and Tradeoff
  Analysis}.
\newblock {\em IEEE/ACM ToN}, 22(2), 2014.

\bibitem{solensky}
F.~Solensky.
\newblock {Continued Internet Growth}.
\newblock In {\em IETF 18}, pages 59--61, 1990.
\newblock \url{http://www.ietf.org/proceedings/18.pdf}.

\bibitem{irr50percent}
Andree Toonk.
\newblock {How accurate are the Internet Route Registries (IRR). (2009-03-28)}.
\newblock
  \url{http://www.bgpmon.net/how-accurate-are-the-internet-route-registries-irr/}.

\bibitem{bgpmon_rpki}
Andree Toonk.
\newblock {Securing BGP routing with RPKI and ROA's. (2011-01-19)}.
\newblock \url{http://www.bgpmon.net/securing-bgp-routing-with-rpki-and-roas/}.

\bibitem{capturingghosts}
S.~Zander, L.~Andrew, and G.~Armitage.
\newblock {Capturing Ghosts: Predicting the Used IPv4 Space by Inferring
  Unobserved Addresses}.
\newblock In {\em ACM IMC}, 2014.

\end{thebibliography}
\end{document}